\documentclass[12pt]{article} 

\usepackage{geometry} 
\usepackage{color}
\usepackage{soul}
\usepackage{graphicx} 

\usepackage{float} 
\usepackage{wrapfig} 

\usepackage{lipsum} 

\graphicspath{{./Pictures/}} 

\begin{document}


\begin{titlepage}

\newcommand{\HRule}{\rule{\linewidth}{0.5mm}} 

\center 


\HRule \\[0.4cm]
{ \huge \bfseries A Roadmap for Canadian Submillimetre Astronomy}\\[0.4cm] 
\HRule \\[1.5cm]

\begin{minipage}{0.8\textwidth}
\begin{center}\large
\emph{Authors:}\\
 \textsc{Tracy Webb}, \textsc{Scott Chapman}, \\
 \textsc{James Di~Francesco}, \textsc{Brenda Matthews}, \\
 \textsc{Norm Murray}, \textsc{Douglas Scott},  \\
 \textsc{Christine Wilson}  
\end{center}
\end{minipage}\\[4cm]
~


{\large \today}\\[3cm] 

\vfill 

\end{titlepage}


\tableofcontents 

\newpage 




\section{Executive Summary}

Submillimetre/millimetre (SMM) wavelengths are a key resource in  the study of our cosmic origins.  From the Cosmic 
Microwave Background through starburst galaxies to debris disks, submillimetre observations probe the formation
of structure in the universe, on all scales.   What was once seen as a niche science with limited technological capabilities 
has now become a mature field.  SMM  observations are now a crucial component of essentially all sub-areas of astronomy,
and are gathered by a suite of facilities with complementary strengths.

Canada holds a long history of excellence and expertise in SMM astronomy, supported not only through continued access
to cutting-edge facilities but by strong involvement at the end-to-end science level. The ALMA facility is still under 
construction but is already the most powerful SMM interferometer in the world, and Canadian scientists are competing 
strongly for ALMA time through a growing user community (doubling from 23 individuals in Cycle 0 to 55 in Cycle 1). The 
expected withdrawal of Canada from the James Clerk Maxwell Telescope, and the current lack of funding for its obvious 
successor CCAT, leaves the Canadian astronomy community at a critical cross-roads.  We must continue to invest in SMM 
facilities at a level that allows Canadians to maintain our position as a world-leader in an important area of science.  
Historically, Canadians have enjoyed access to the largest, best equipped SMM single dish in the world.  Without such 
continued access, Canadian scientists will lack a key facility for the study of the SMM universe -- one fundamental to their 
future success with ALMA.

This document grew from the discussions about the future of SMM astronomy in Canada during a meeting of interested 
individuals in the SMM community in early 2012.   Here we summarize the observational landscape of SMM astronomy in 
Canada from now to $\sim$2020.  The plans and priorities of the SMM community have been discussed extensively in the 
reports of the Canadian astronomy Long Range Plan panels.  A number of recent changes, particularly the demonstrable 
success of SCUBA- 2 that led to a resurgence in the JCMT user community (tripling from 14 individuals on PI proposals 
in semester 11B to 49 in semester 12B),  necessitate a revision.  To maintain the spirit and primary recommendations of
the LRP reports,  Canada must remain flexible and ready to respond appropriately to current realities and new
opportunities.

We argue that because of Canada's substantial investment in ALMA and numerous PI-led SMM experiments,
continued involvement in a large single-dish facility is crucial.  In particular, we recommend: \\

 (i) an extension of Canadian participation in the JCMT until {\it at least} the unique JCMT Legacy Survey program
is able to realize the full scientific potential provided by the world-leading SCUBA-2 instrument; \\

 (ii) involvement of the entire Canadian community in CCAT,  with a large enough share in the partnership that
Canadian astronomers can significantly participate at all levels of the facility (decisions, construction, and science). \\

We further recommend continued participation in ALMA development, involvement in many focused PI-led
SMM experiments, and partnership in SPICA.  We close with an outline of the expected costs of these
recommendations and options for funding. 

\newpage
\section{Submillimetre/millimetre Astronomy}

Submillimetre/millimetre (SMM)  wavelengths encompass the wide
range between $\sim$200 $\mu$m and $\sim$10 mm.   Over the last three decades, SMM astronomy has grown from a field
studying small numbers of near-by objects, to a science capable of undertaking comprehensive investigations
of all aspects of the universe, with data quality comparable to what was traditionally only reached with  optical wavelengths.  
The strength and importance of SMM astronomy is extensively outlined elsewhere (such as the SCUBA-2 Legacy Survey Proposals or the CCAT Science Documents),
but for completeness we highlight some important aspects here.

SMM radiation can be broadly characterized as the light of our cosmic origins. At its lowest frequencies, it probes
the very nature of the universe through the primary Cosmic Microwave Background, and the additional imprints of  foreground
matter in the form of secondary anisotropies (galaxy clusters, gravitational lensing etc).  Locally, SMM observations are the work-horses
of astronomers seeking to understand the very basics of planet and star formation. In between these two extremes, SMM measurements are
 diagnostics of the   most energetic and dynamic phases of galaxy formation and evolution - the epochs of stellar mass assembly and supermassive
 black hole growth. 

As SMM radiation is sandwiched between radio and infrared frequencies, specialized
instruments incorporating both optical and radio techniques are required.    Moreover, the
wavelength range actually observable depends strongly on atmospheric conditions,
typically precipitable water vapour content, and some SMM wavelengths
are simply inaccessible from the ground from even the driest terrestrial sites. This has
resulted in a number of technical challenges which traditionally limited the field, but
are now being routinely overcome.

SMM measurements can be generally divided into three distinct categories which provide complementary information. 
We briefly outline these below and highlight the primary scientific uses of each.
 
 \vskip 0.75cm
{\it Continuum Emission}:
SMM wavelength continuum observations  trace extended emission
from dust particles within the interstellar medium (ISM).  This dust is warmed
by nearby hot sources (e.g., young stars) and the interstellar radiation field,
and they cool through  thermal emission.  Therefore, this emission traces
the thermal balance in the ISM.  In addition, the emission is generally optically
thin, so the observed intensity is proportional to the column density and temperature
of the emitting dust.  Hence, SMM continuum emission is an excellent tracer of
mass in the ISM.  For example, protostellar or debris disks contain dust warmed
by central stellar objects, and these objects glow brightly at SMM wavelengths.
Moreover, the dense predecessors to star formation, dense cores, can be well
traced by the thermal SMM emission of its dust.  SMM continuum emission can
also trace galactic structure, such as the dusty spiral arms of galaxies or the interaction
zones of colliding galaxies.  Moreover, the earliest galaxies are heavily enshrouded
by dust, and SMM continuum emission can trace the vigorous star formation rates
within high redshift galaxies.  Beyond thermal emission from dust, SMM continuum
emission can also trace the cosmic microwave background (CMB) and specifically
its anisotropies that trace conditions in the early universe.  Furthermore, localized
deficits or enhancements in the CMB toward galaxy clusters can be caused by the
Sunyaev-Zel'dovich Effect, and galaxy cluster masses can be so probed using
SMM continuum observations.  At the longer SMM wavelengths, i.e., $\geq$ 1 mm,
continuum emission may also include significant amounts of free-free radiation
from ionized material, e.g., shocked gas in protostellar outflows. \\

{\it Transitions from Molecular or Atomic Gas}:
Spectral lines are also found at SMM wavelengths, and these provide very
important views of galactic and extragalactic environments.  For example, the
SMM range includes the rest wavelengths of numerous molecular rotational
transitions that are excited by conditions found in the ISM (see Figure 1).  Since
profiles of lines inherently trace velocity structure along a given direction, lines
can be used to probe the kinematics and dynamics of gas within the ISM.  Moreover,
given the strong dependence of transitions on specific excitation conditions, the
lines also trace such conditions throughout the ISM.  For example, observations
of rotational line emission from CO (or its heavier isotopologues) trace the cold
components of disks, as well as structure of molecular clouds and their dense
cores throughout the Galaxy.  Furthermore, molecular lines can trace directly
the physical conditions and kinematics of both nearby galaxies and distant
objects at high redshift.  (In addition, the latter objects can have key atomic
lines like the ones of [OI] at 63 $\mu$m, [CII] at 158 $\mu$m, or [NII] at 205
$\mu$m, shifted into the SMM range, allowing expanded probes of their ISMs.)
The relative brightnesses of the transitions of several molecular species is a
also key probe into the chemistry of the ISM, with the relative abundance of
various molecules playing an important role in its thermal balance. \\

\begin{figure}[H] 
\center{\includegraphics[width=0.8\linewidth]{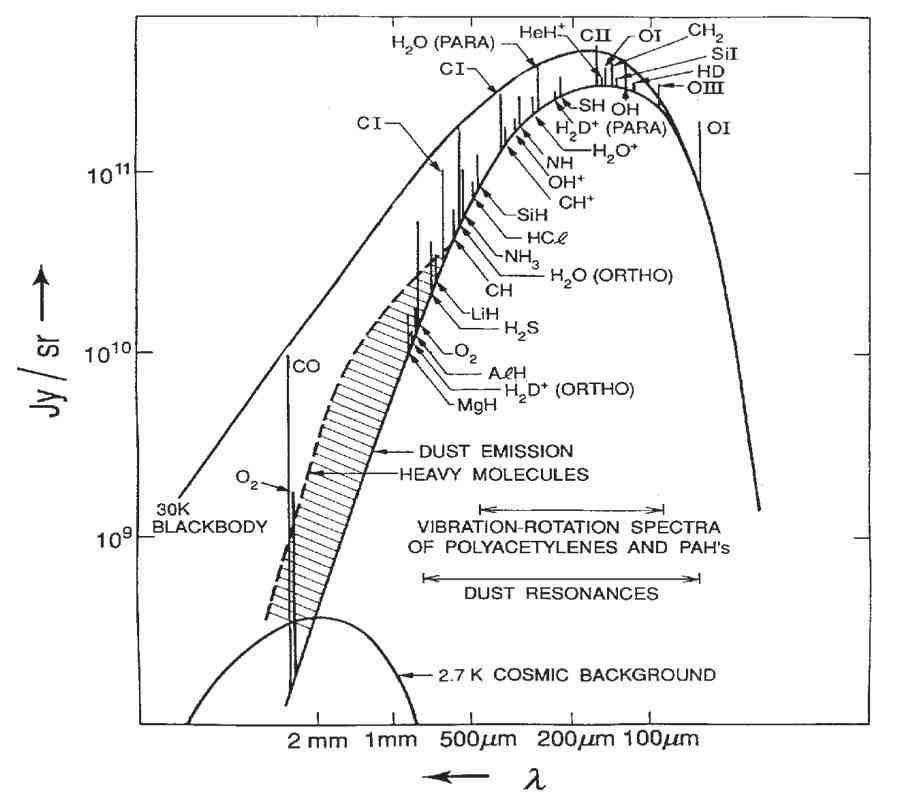}}
\caption{Schematic representation of some of the spectral content at SMM
wavelengths toward an interstellar cloud (Phillips \& Keene 1992).}
\label{fig:speciation}
\end{figure}

{\it Polarization from Continuum or Line Emission}:
SMM continuum emission and some lines can also trace magnetic fields in
the ISM.  The roles magnetic fields play in the dynamical evolution of the ISM
are poorly constrained and SMM data remain among the most effective means
to trace them.  For continuum observations, thermal emission from dust can be
measurably (linearly) polarized when magnetic fields preferentially align the orientations
of the emitting grains. These data can reveal the magnetic field strengths in the
plane of the sky.  In addition, polarization of the CMB continuum emission can
be used to probe the imprint of gravitational waves during the earliest moments
of the universe.  For certain line observations, the emission from some molecules
(e.g., CCS, CN) can be measurably split and (circularly) polarized due to the Zeeman
effect, i.e., the rotational energy levels are subtly shifted in presence of magnetic
fields.  These data probe the magnetic field strengths along the line of sight.
Importantly, SMM continuum and line observations trace denser and more
compact structures than are possible to observe using lower-frequency
radio observations, e.g., using HI or OH lines.

\vskip 0.75cm

As with any other wavelength regime, SMM astronomy requires different
facilities and instruments to achieve different scientific goals. No one instrument
or facility can meet the needs of all the lines of inquiry available.  Indeed, numerous
observatories with diverse instrumentation have been built over the last three decades
to exploit the key information residing within SMM wavelengths.  Single-dish
facilities like the James Clerk Maxwell Telescope (JCMT) and
the Green Bank Telescope (GBT) provide the larger-scale view of the SMM universe.
(Of special note are airborne and spaceborne observatories like the Balloon-borne
Large Aperture Submillimetre Telescope (BLAST), its successor BLAST-Pol, and the
Herschel Space Observatory, which trace submm wavelengths unobservable from the
ground.) Such facilities, however, are diffraction limited in this wavelength range, and
achieving high-resolution images (e.g, subarcsecond) requires interferometric techniques.
Such observatories, including now the Atacama Large Millimetre/submillimetre
Array (ALMA) and the Karl G.~Jansky Very Large Array (JVLA), all serve at present to provide the
smaller-scale view.   By necessity, however, these data come with intrinsic trade-offs
in sensitivity to total flux and larger-scale structures.  The complementary nature of
single-dish and interferometers underscores the need for access to both types of 
facility  (Indeed, in certain cases, data from each  may be combined
to provide highly detailed images with information on a wide range of spatial scales.)

%
%
 
With over 25 years of access to the JCMT, Canadian astronomers have developed
a worldwide reputation as leaders in SMM research.  They have recently had access
to JCMT, GBT, JVLA, ALMA, BLAST, Planck, and Herschel, enabling high-impact 
astronomical research at the world's top SMM observatories.  If Canada is to maintain
its leadership in astronomical research, however, it cannot do so without continued
access to, and support of, the best SMM facilities and instrumentation available worldwide.
In the view of the SMM community, this includes BOTH single-dish and interferometric
facilities.  In this document, we survey the present and future for Canadian SMM facilities,
to provide a vision of continued leadership (i.e., a ``roadmap") for Canada in this very
important wavelength regime.



\newpage
\section{SMM in the Context of Canadian LRPs}


The Long Range Plan (LRP) series has served Canada well by using community
input to outline the priorities and
directions of Canadian astronomy for the following 10-15 years.  The most recent version, LRP2010,
recommended Canadian investment in 13 new facilities, with a total estimated cost of \$550M  over ten
years.  The priority rankings of these facilities were organized according to small, medium, and large cost
scales, and between space- and ground-based projects.

In LRP2010, the highest priority identified for a large-scale, ground-based project was 
Canadian participation in
the optical Thirty Meter Telescope (TMT) project.  While final construction budgets are not yet known, TMT is
expected to cost at least $\sim$\$1.2B, with Canada contributing 25\% (or \$305M).   Once TMT is
constructed by the end of the decade, the Square Kilometer Array (SKA) will become Canada's highest ground-based priority.  In
the medium-scale category, LRP2010 recommended three projects, at \$15M each, one of which -- CCAT -- is a new SMM facility.  The other two include the radio
Canadian Hydrogen Intensity Mapping Experiment (CHIME - now funded through a CFI grant)  and upgraded instruments on the optical
Canada-France-Hawaii Telescope (CFHT).  The top two small-scale projects ($<$ \$5M) are an
arctic-based optical facility and a study for the Next Generation CFHT (ngCFHT). 

For space-based projects, LRP2010 gave highest priority to Canadian participation in a large-scale
dark energy mission of some kind, such as the European Space Agency (ESA) Euclid project, at \$100M.
Recommended medium-scale, space-based projects included participations in the NASA-ESA International
X-ray Observatory (IXO) and the follow-up to Herschel, the Japanese-led far-infrared/submm Space Infra-Red
Telescope for Cosmology and Astrophysics (SPICA). 

LRPs summarize a set of recommendations and guidelines, identified at the beginning of a decade,
based on scientific aims.  As noted carefully in the LRP2010 report, however, {\it ``these are provisional
rankings and must be qualified: they are on the basis of science promise and/or long-term potential impact only.
All of these projects lack a thorough feasibility study, technical review, and cost analysis..."} 
As new information or opportunities arise, a flexible plan is needed to  
maximize the scientific return of our current investments.   For example, although LRP2010 reiterated
phasing out Canadian participation in the JCMT as funding for ALMA ramps up, this recommendation
was crucially tied to the then-unknown performance of the flagship SCUBA-2 instrument on JCMT.  The report to the LRP2010 panel by the Canadian Astronomical Society's Ground Based Astronomy
Committee (GAC), recommended the transition from the JCMT to ALMA should be done in a timely manner. Since
the publication of LRP2010, SCUBA-2 performance has been determined to be of very high quality, and
the subsequent response by the Canadian community to use SCUBA-2 has been overwhelmingly
positive (see \S4.1 below).  As such  the JCMT remains an extremely relevant facility today.

In parallel, Canadian interest in CCAT remains strong, following the recommendation by LRP2010.  Indeed,
present Canadian participation in CCAT is due to a strong grass-roots effort to raise the initial funds at the university level.
The funding source for the minimum  contribution to CCAT (\$20M - as required by the telescope consortium) has not yet been identified, though
the project is moving forward on schedule and predicted to reach first light in 2018. 
Similarly, SPICA, though targeted for launch in 2022, has not yet been approved by the Japanese Space
Agency (JAXA).

\newpage
\section{The Submillimetre/Millimetre User \\
 Community in Canada}

The SMM facility user-community is a significant demographic within the larger astronomy community in Canada.  It is difficult to objectively and uniquely quantify the size of the community since it encompasses observational astronomers working in multiple wavelength regimes, theorists, and experimentalists building PI-lead instruments.  We therefore turn to the user rates of Canada's forefront facilities as an objective, if incomplete, metric.  Table 1 shows the involvement level of Canadian astronomers in ALMAÕs Cycle 0 (mid-2011) and Cycle 1 (mid-2012) calls for proposals.
The fraction of allocated programs led by Canadian PIs rose from 2.7\% to 3.1\% between Cycle 0 and Cycle 1, but more significantly, the number of Canadian astronomers (i.e., those at Canadian institutions) involved in ALMA proposals worldwide doubled from 23 to 55 individuals, likely arising from the improved capabilities of ALMA from one cycle to the next.

\begin{table}[htdp]
\caption{Canadian ALMA User Statistics}
\begin{center}
\begin{tabular}{cccccc}
\hline
Cycle & Total  & Canadian& Total & Allocated to & Success   \\
& Worldwide & Lead & Canadians & Canadian & Rate of \\
 & Proposals & Proposals & Involved &  PIs & Canadian PIs \\
\hline
0 & 924 & 24 (2.6\%) &23 & 3 of 112 (2.7\%) & 3 of 24 (12.5\%) \\
1 & 1133 & 26 (2.3\%) & 55 & 6 of 196 (3.1\%) & 6 of 26 (23\%) \\
\hline
\end{tabular}
\end{center}
\end{table}

A comparison of the numbers of unique Canadian users applying for PI time on Canada's three major off-shore 
single-aperture facilities shows that the number of JCMT users exhibited significant fluctuation over the four
semesters tracked (Table 2). The large increase in unique users in semester 12A can be attributed to the
availability of the SCUBA-2 instrument on the telescope.  The number of users applying for time in 12B is three
times the number for 11B when SCUBA-2 was not offered (and is larger than for the CFHT in the same semester).
Note, however, that these numbers do not take into account users of JCMT through the JCMT Legacy Survey
who are not involved in PI programs in a given semester.

\begin{table}[htdp]
\caption{Numbers of Unique Canadian Investigators on Proposals to Canada's Three Major Single-Aperture Off-Shore Facilities}
\begin{center}
\begin{tabular}{cccccc}
\hline
  Facility  & 11A & 11B & 12A & 12B & 13A \\
\hline
JCMT & 15 & 14 & 31 & 49 & 48  \\
CFHT & 53 & 39 & 44 & 47 & 54 \\
Gemini & 48 & 58 & 46 & 71 & 85 \\
\hline
\end{tabular}
\end{center}
\end{table}
~

\bigskip
\begin{figure}[H] 
\center{\includegraphics[width=1.0\linewidth]{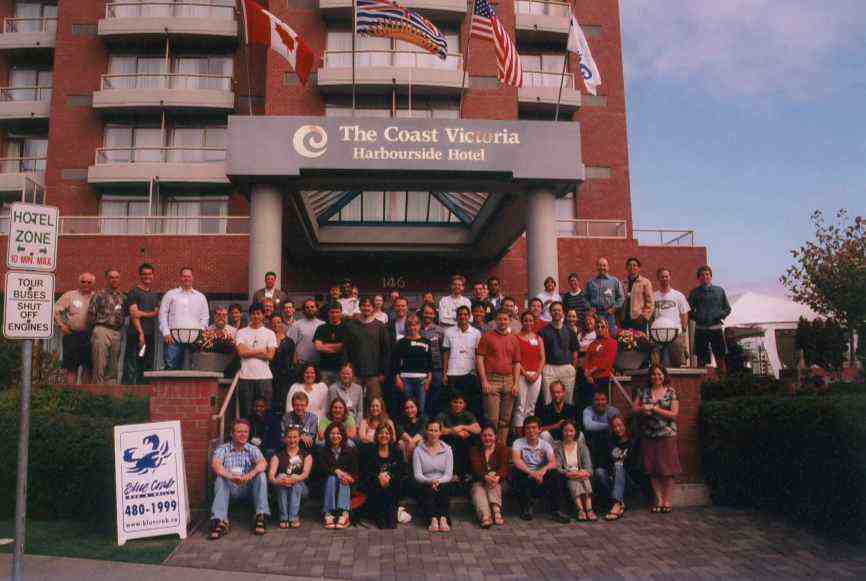}}
\caption{Participants of the NRC (Sub)Millimetre Observing Techniques
Summer Workshop held July 2006 in Victoria, BC.}
\label{fig:speciation}
\end{figure}

\newpage
\section{The Present SMM Landscape in Canada} 

\subsection{The Atacama Large Millimetre/submillimetre Array}


The Atacama Large Millimeter/submillimeter Array (ALMA) is the first billion-dollar ground-based
astronomical telescope.  It is an SMM interferometer operating on the Atacama desert of northern
Chile.  ALMA was identified as Canada's first priority for ground-based facilities in the LRP2000
document, and Canada formally joined the project in 2003 through the North American Partnership
for Radio Astronomy (NAPRA).  ALMA is presently a collaboration between three partner regions,
North America (US, Canada, and Taiwan), Europe (the member states of ESO), and East Asia (Japan
and Taiwan, again) and the host country, the Republic of Chile.  After years of planning and construction,
ALMA is finally nearing completion.  Even in its present form, however, ALMA is already the most
sensitive SMM interferometer on the planet.  It will remain Canada's most powerful shorter wavelength
SMM facility for many years.  

\bigskip
\begin{figure}[H] 
\center{\includegraphics[width=0.8\linewidth]{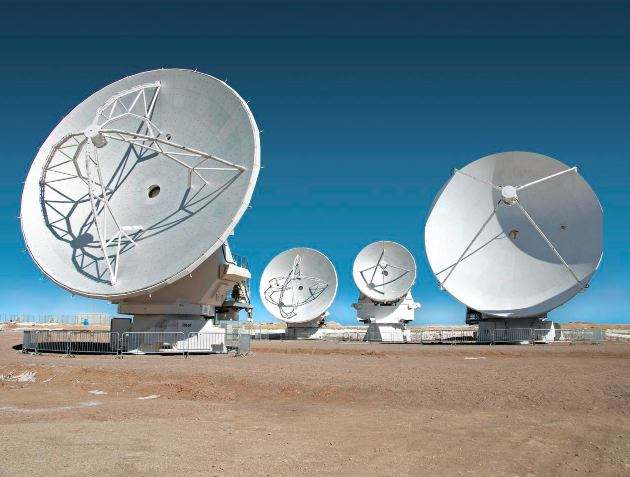}}
\caption{Four models of ALMA antennas on the Chajnantor plateau. (Credit: ALMA (ESO/NAOJ/NRAO),
W. Garnier (ALMA)}
\label{fig:speciation}
\end{figure}

When complete, ALMA will consist of fifty 12-m antennas, as well as twelve 7-m antennas and
four 12-m antennas specifically designed for compact configuration and total power observing,
respectively.  Each antenna is equipped with a suite of single-pixel receivers that operate over
specific wavelength bands (see Table 3).  ALMA observations are made with one receiver per
telescope, though in principle ALMA could be divided into four subarrays of telescopes using
different receivers simultaneously. Received signals are sent to one of two correlators that
process the data into continuum data or spectral line data cubes.  

\begin{table}[htdp]
\caption{ALMA Receiver Bands}
\begin{center}
\begin{tabular}{ccc}
\hline
Band & $\lambda\lambda$ (mm) & Availability \\
\hline
1 & 6 -- 8.5 & under development \\
2 & 3.3 -- 4.5 & under development \\
3 & 2.6 -- 3.6 & Cycle 0 onwards \\
4 & 1.8 -- 2.4 & Cycle 2 onwards \\
5 & 1.4 -- 1.8 & under development \\
6 & 1.1 -- 1.4 & Cycle 0 onwards \\
7 & 0.8 -- 1.1 & Cycle 0 onwards \\
8 & 0.6 -- 0.8 & Cycle 2 onwards \\
9 & 0.4 -- 0.5 & Cycle 0 onwards \\
10 & 0.3 -- 0.4 & Cycle 2 onwards  \\
\hline
\end{tabular}
\end{center}
\end{table}




ALMA has already had two ``Early Science" proposal calls to use its Band 3, 6, 7, and 9
receivers.  Cycle 0 observations, using a minimum of sixteen 12-m antennas, took place throughout 2012.   Cycle 1 observations, using a minimum of thirty-two 12-m
antennas (and, if needed, nine 7-m compact configuration antenna and two 12-m total
power antennas) began in early 2013.   During the Early Science phase, scientific time at
the telescope is still being tensioned with ongoing construction, and all observations are only expected 
on a best-effort basis.  Cycle 2 observations are expected to be part of the Full Science
phase, using all 66 antennas and Bands 3, 4, 6, 7, 8, 9, and 10, and these will likely begin
in 2014.  (NB: Bands 1, 2, and 5 are under development.)


Through NAPRA,
Canadian astronomers have access to the 33.75\% of ALMA time available to North American
astronomers,
with no specific percentage of time defined as Canadian.  In comparison, Canada provides
7.25\% of the cost of North American ALMA operations.  

Demand for ALMA time has been extraordinarily high.  For Cycle 0 (mid-2011), 924 proposals
in total were submitted for $\sim$700 hours of observing time, with 24 led by PIs at a Canadian
institution.  
Only the top 112 proposals ($\sim$12\%) received ``high priority" status, including
three by Canadian PIs.  For Cycle 1 (mid-2012), 1133 proposals were submitted for $\sim$800
hours of observing time, with 26 led by Canadian PIs.  
(Including co-Is, 55 individuals from 14 Canadian institutions were involved in Cycle 1 proposals.)
Only the top 196 proposals ($\sim$17\%) received ``high priority" status, including six by Canadian PIs.
These numbers show that Canadians have to date competed successfully for ALMA time at levels just
above that of our monetary contribution.  Most importantly, over two proposal cycles, the number of
successful proposals from PIs at Canadian institutions has increased.  Cycle 2 proposals are expected
to be due in  September 2013.

\subsection{James Clerk Maxwell Telescope}
The James Clerk Maxwell Telescope (JCMT), located on the dry summit
of Mauna Kea, has been Canada's primary access to SMM wavelengths for
25 years.  At 15-m diameter, JCMT remains the world's largest present single-dish
SMM facility.  Historically, the JCMT has been a partnership between the UK (55\%),
Canada (25\%) and the Netherlands (20\%). (The University of Hawaii 
receives 10\% of time as the host institution.)  The JCMT has been
managed by the Joint Astronomy Centre in Hilo, Hawaii together with the
United Kingdom InfraRed Telescope (UKIRT).  

\bigskip
\begin{figure}[H] 
\center{\includegraphics[width=0.8\linewidth]{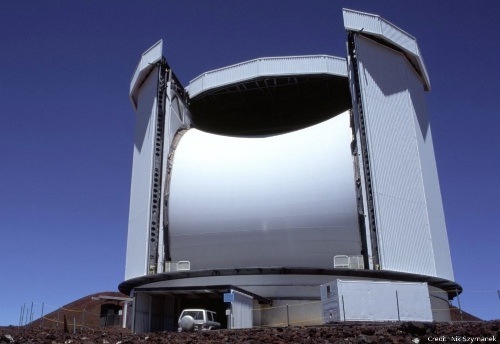}}
\caption{The James Clerk Maxwell Telescope. (Credit: Nik Szymanek) }
\label{fig:speciation}
\end{figure}

The instrumentation suite at JCMT has made a very significant scientific
impact over its lifetime, mostly due to SCUBA, the first submillimetre bolometric
``camera."  In the early 2000s when SCUBA was in its prime, JCMT was
considered top of its class. JCMT's instrumentation has all been updated or
completely replaced in the last 5-10 years and remains relevant and cutting-edge.
The current suite consists of:

~

$\bullet${{\bf ~SCUBA-2} - an $\sim$8$^{\prime}$$\times$8$^{\prime}$-wide
850 $\mu$m and 450 $\mu$m bolometric camera};

$\bullet${{\bf ~HARP} - a $\sim$2$^{\prime}$$\times$2$^{\prime}$-wide 4 $\times$
4 325-375 GHz heterodyne receiver array};

$\bullet${{\bf ~RxW} - two single-pixel receivers at 315-375 GHz and 630-710 GHz}; and

$\bullet${{\bf ~RxA} - a single-pixel receiver at 211-272 GHz.}

~

\noindent
The fabrication of SCUBA-2  was a collaborative effort between the UK, JAC and several Canadian
partners. The Canadian effort involved the warm electronics and data reduction software and was funded largely through a grant from the Canadian Fund for Innovation.  Now fully on-line, it produces high-quality continuum images of unprecedented
sensitivity at 850 $\mu$m and 450 $\mu$m. HARP uniquely produces
stunning wide-field data cubes of similar resolution to SCUBA-2 at 850
$\mu$m.  The remaining receivers, RxW and RxA, are older, but they still
provide very good scientific return when conditions warrant.  Signals
received by all three heterodyne JCMT instruments are fed to ACSIS,
the digital autocorrelator spectrograph built in Canada by NRC-HIA.

Currently, JCMT users access the telescope through the JCMT Legacy
Survey (65\% of time) and PI time (35\%).  The strong emphasis on Legacy
science is viewed favourably by the wider community.   Ample amounts of
PI time, however, allow for the follow up of Legacy Survey discoveries,
exploration of science not included in the Legacy Surveys, and utilization
of SCUBA-2's ancillary instruments: POL-2, a polarimeter, and FTS-2, a
Fourier Transform spectrometer, both of which were built by Canada
(Montreal and Lethbridge, respectively).  In the following, we explore
further both modes of JCMT access.

\subsubsection{The JCMT Legacy Survey}

With its latest wide-field instrumentation, HARP and SCUBA-2, the JCMT is able
for the first time to conduct very large surveys.  Recognizing this opportunity, a
new time allocation model was adopted, whereby a majority of UK, Canada, and
Dutch observing time was set aside for the large tri-national JCMT Legacy Survey (JLS).
Seven components of the JLS were approved in 2005 after extensive peer review,
including cosmology (CLS), nearby galaxies (NGS), the Galactic plane (JPS), a
spectral line survey of star-forming objects (SLS), nearby star formation (GBS),
debris disks around nearby stars (SONS), and an ``all-sky" survey (SASSy).
JLS HARP observing began in late 2007 and is effectively complete.  JLS
SCUBA-2 observing began in late 2011 and is ongoing.

Canadians are heavily invested in all components of the JLS.  Canadian researchers have
leadership roles (Co-Principal-Investigators) on all components and the fraction of Canadian
participants on each is 20-40\%.  

Most of the JLS time requested was intended for SCUBA-2 observations.  In fall
2011, the JLS components needing SCUBA-2 data submitted re-scoped proposals
for a new round of peer review, using the actual on-sky performance of SCUBA-2.  This process was driven
by the limited time available on the telescope, given the impending end of the agreement to operate the JCMT, and not by a 
lack of scientific interest; completing the peer-reviewed goals of the original JLS would take several
times longer than the time now available. The proposals were granted
a total of 3490 hours (291 nights) over two years, divided among various weather
bands (see Table 4).  For  comparison, the previous JLS allocation
of time with HARP was 962 hours. 

\begin{table}[htdp]
\caption{SCUBA-2 Hour Allocations to JLS Components by Weather Band}
\begin{center}
\begin{tabular}{ccccc}
\hline
Component  & Band 1 & Band 2 & Band 3 & Band 4 \\ \hline
CLS & 629 & 695 & 454 &  \\ 
GBS & 70 & 342 & & \\ 
JPS & & & 450 & \\
NGLS & & 100 & &  \\
SASSy & & & & 480 \\
SONS & &135 & 135 & \\ \hline
\end{tabular}
\end{center}
\label{default}
\end{table}%

\subsubsection{PI-led Projects}

Significant JCMT time remains for PI-led projects. Historically, Canadian use
of the JCMT was oversubscribed by factors of $\sim$3 in the early- to mid-2000's.
In the period between SCUBA and SCUBA-2, the
oversubscription rate peaked at $\sim$5 in semester 06B but declined to ~1.0 in
semesters 11A and 11B.  (Note, however, that no continuum instrument was available
at that time and many members of the Canadian community were engaged in HARP
aspects of three JLS components.)  Now with access to SCUBA-2, Canadian PI interest
in JCMT has returned, though amounts of PI time available are reduced due to the
large JLS allocation.  Indeed, when the JLS components were re-scoped in fall 2011,
some aspects (e.g., certain sky coverages) became impossible to complete and
this time was made available back to the community.  In semesters 12A-13A, the
oversubscription rate was on average 5.9\footnote{The oversubscription rates at
JCMT for semesters 12A, 12B, and 13A were 5.6, 8.1, and 3.9, respectively.}.
{\it These numbers clearly indicate the strong interest in SCUBA-2,  the other
instrumentation, and JCMT in general in PI-led science.}

Though the JLS was defined to address a wide range of astrophysical
phenomena, it is critical for the community to have access to the JCMT for
smaller, focused projects.  For example, such small projects can respond to
scientific developments occurring since the JLS was defined (or re-scoped).
Hence, a healthy share of JCMT time should remain for PI-led projects.  

%

\subsection{Herschel Space Observatory}

The Herschel Space Observatory deserves special mention in the current
Canadian SMM landscape.  This ESA-led mission consists of a 3.5-m SMM
telescope located at the Sun-Earth L2 point.  Herschel was designed to observe
SMM wavelength ranges impossible to observe from the ground due to strong
atmospheric absorption.  Herschel has three instruments: 

~

$\bullet${\bf ~HIFI} - a heterodyne spectrometer that can observe frequencies over
ranges of 480-1250 GHz and 1410-1910 GHz;

$\bullet${\bf ~SPIRE} - a photometric array with FTS capabilities that can observe
at 250 $\mu$m, 350 $\mu$m, and 500 $\mu$m; and 

$\bullet${\bf ~PACS} - a photometric array with grating spectrometer capabilities that
can observe at 70 $\mu$m, 100 $\mu$m, and 160 $\mu$m. 

~

\noindent
Canada, through the Canadian
Space Agency (CSA), contributed to the construction of the first two of these
instruments, through support of efforts at U. Waterloo (M. Fich) and U. Lethbridge
(D. Naylor).  Herschel was launched successfully by ESA in May 2009 and is
expected to run out of cryogens and become inoperable in February-March
2013. 

Herschel's time was divided between $\sim$33\% Guaranteed Time provided
to the consortia that built its three instruments and  $\sim$66\% Open Time to
the world community.  Both allocations were further divided into Key Projects
requiring $>$100 hours and smaller Regular Projects.  Canadians participated
in Herschel projects within the Guaranteed Time and Open Time allocations,
at both the Key and Regular Project levels.  (The former projects were made 
possible through agreements between CSA and ESA.)

\bigskip
\begin{figure}[H] 
\center{\includegraphics[width=0.8\linewidth]{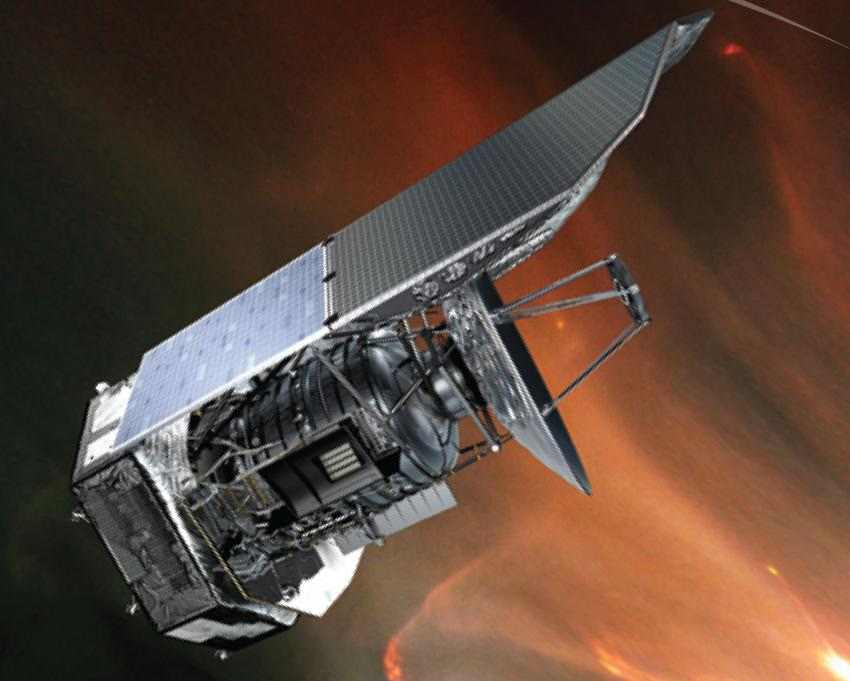}}
\caption{The Herschel Space Observatory. (Credit: ESA / AOES Medialab, background:
Hubble Space Telescope image (NASA/ESA/STScI)}
\label{fig:speciation}
\end{figure}

Herschel's scientific output has been impressive, producing wide
far-infrared and submm continuum maps at resolutions similar to those
obtained previously from the ground and spectral line data cubes at THz 
frequencies of unparalleled sensitivity.  Exploitation of Herschel data by
Canadians is ongoing but all Herschel data will reside for all future users 
worldwide in raw and pipeline-reduced forms in an ESA-managed archive.
In some ways, Herschel data can be used as a pathfinder for future ALMA
observations but the two facilities only overlap at the shortest wavelengths
ALMA can observe, Bands 9 and 10.  Given the relatively small size of
Herschel's aperture, its data cannot be combined directly with ALMA data
at overlap wavelengths.  No other SMM space-based observatories have
been planned worldwide besides the not-yet-approved, Japanese-led SPICA
mission slated to launch in 2022 (see \S5.5).

\bigskip
\begin{figure}[H] 
\center{\includegraphics[width=0.8\linewidth]{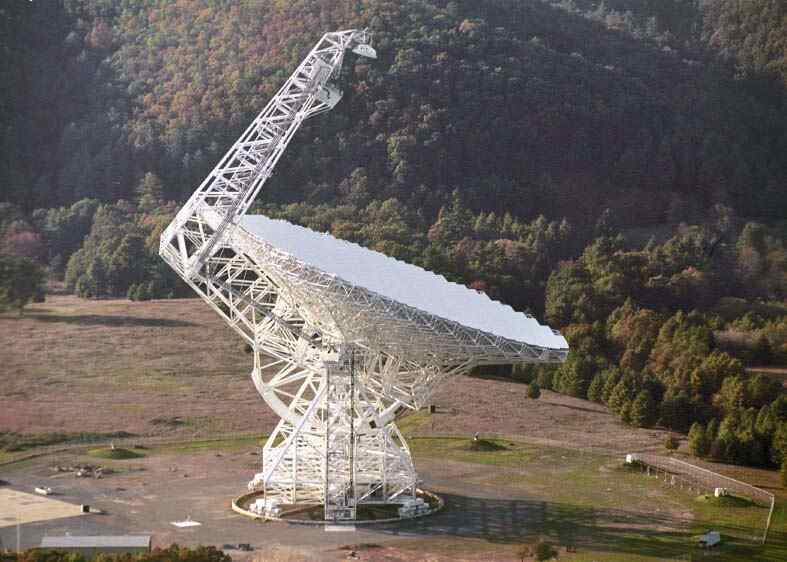}}
\caption{The NRAO Robert C. Byrd Green Bank Telescope. (Image courtesy of NRAO/AUI)}
\label{fig:speciation}
\end{figure}

\subsection{Green Bank Telescope and Jansky Very Large Array}

The Robert C. Byrd Green Band Telescope (GBT) and Karl G. Jansky Very Large
Array (JVLA) are longer-wavelength SMM/radio facilities operated by NRAO in the
United States.  The GBT is a 100-m diameter fully steerable single-dish antenna 
located at Green Bank, West Virginia, equipped with three single-pixel receivers
that reach different bands over the longest wavelengths of the SMM range, 3 mm
and 7-10 mm, and MUSTANG, a bolometer array that can observe continuum
emission at $\sim$3 mm.  The Jansky Very Large Array (JVLA) is an expansive
retrofit of the Very Large Array interferometer located near Socorro, New Mexico.  As
with the VLA before, the JVLA consists of twenty-seven 25-m diameter antennas
with each antenna equipped with a suite of heterodyne receivers that include
wavelengths of 7-10 mm.  Indeed, the GBT's and JVLA's abilities to observe
wavelengths of $\sim$3 mm and 7-10 mm, and Canada's access to them through
the NAPRA, demand their inclusion here.  Though the GBT has an impressive aperture,
its site is not ideal for mm observations.  It can be sensitive when conditions are
right but its efficiency is much greater at longer (radio) wavelengths.  The JVLA is
a big improvement over the older VLA in terms of aperture efficiencies and
wavelength coverage.  The heart of the JVLA's profound sensitivity improvement
is, however, its new wide-band WIDAR correlator, which was built by NRC-HIA.
(This contribution enabled Canada in part to join the ALMA project; see below.)  
The JVLA site is better than the GBT site for long-wavelength SMM observations.

\begin{figure}[H] 
\center{\includegraphics[width=0.8\linewidth]{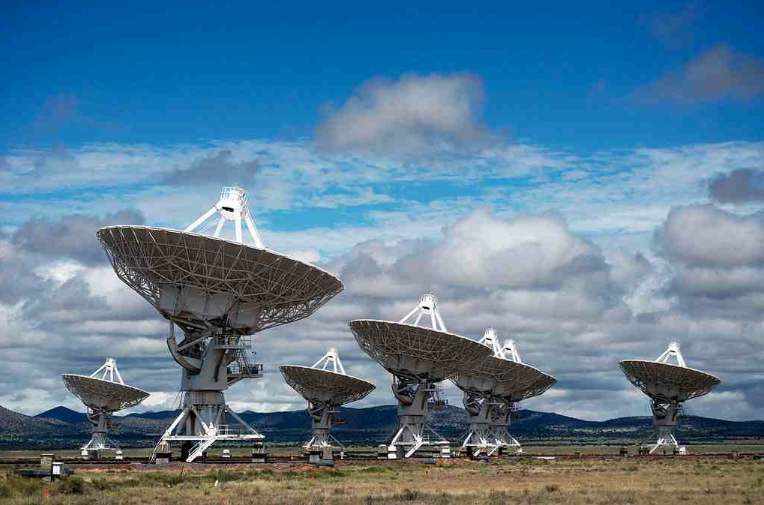}}
\caption{The NRAO Karl G. Jansky Very Large Array. (Image courtesy of NRAO/AUI)}
\label{fig:speciation}
\end{figure}

The GBT presently operates in Full Science mode.  The JVLA is moving from an
Early Science phase to a Full Science facility as its vast number of correlator modes
are commissioned.  Through NAPRA, Canada does not provide funds for the
operations of the JVLA or GBT.  It has no fixed share of GBT or JVLA time and
all Canadian proposals are reviewed in the same manner as all others.  At present,
the GBT and JVLA provide for Canada premier access to high-resolution continuum
and line emission observations at longer SMM wavelengths.  (No data about the
extent of Canadian participation in recent GBT or JVLA proposal rounds have been
compiled for this document.)  The future of the GBT is unclear at present; in August
2012, the US National Science Foundation (NSF) AST Portfolio Review Committee (PRC)
recommended that in the face of flat or declining NSF budgets, the NSF divest itself
from its present support of the GBT so funds can be reallocated to new, high-priority
astronomy projects.  The JVLA's future appears more secure, however; it was not 
targeted for NSF divestment by the PRC.  The JVLA is presently the most sensitive
facility on the planet over the wavelength range of 7-10 mm (and most radio
wavelengths, too).  Canada is presently in a partnership with Taiwan, Japan, Chile,
and the US to develop Band 1 receivers for ALMA, which will enable $\sim$6-8.5
mm observations from the southern hemisphere at sensitivities moderately
improved over those possible with the JVLA.

\begin{table}
\caption{Current PI Experiments involving Canadian Researchers}
\begin{center}
\begin{tabular}{lcccc}
\hline
facility & aperture & location & focus & access \\
\hline
ACT & 6m & Atacama, Chile & CMB / CMB pol & closed \\
APEX-SZ & instrument & -- & SZ effect/CMB & closed \\
BLAST & & balloon/SP & CMB/SF & closed \\
BLAST-Pol & & balloon/SP & SF pol & closed \\
FTS-2 & instrument & JCMT & broad & open \\
HIFI & instrument & Herschel & broad & open \\
POLARBEAR & 4m & Atacama, Chile & CMB pol & closed$^{1}$ \\
POL-2 & instrument & JCMT & broad & open \\
SPT & 10m & South Pole & CMB & closed$^{2}$ \\
SCUBA-2 & instrument & JCMT & broad & open \\
SPIRE-FTS & instrument & Herschel & broad & open \\
\hline
\end{tabular}
\end{center}

$^{1}$ open for non-CMB science

$^{2}$ closed for now

\end{table}

\subsection{Focused PI Experiments}


Though aforementioned facilities are open to the entire Canadian community, some members
have access to smaller SMM telescopes that are typically led by PIs and focused on specific
experiments.  These projects often produce survey-style datasets that address certain goals
but these products may be suitable for a host of other unanticipated uses.  In Table 5, we outline 
those focused PI-led SMM facilities that currently include Canadian participation.  Unless otherwise
noted, these facilities operate as consortia closed to external collaboration, though the data
products likely enter the public domain after some period. 

\begin{figure}[H] 
\center{\includegraphics[width=0.8\linewidth]{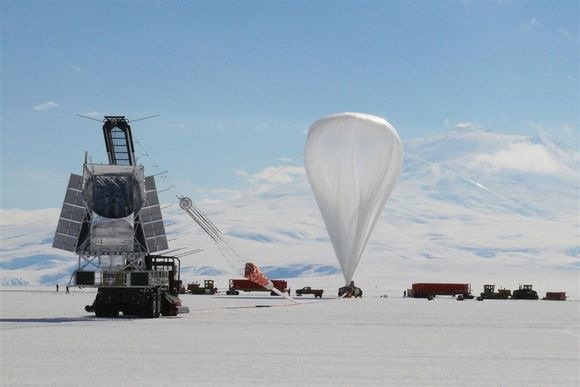}}
\caption{BLAST and its launch balloon in Antarctica. (Credit: M. Halpern)}
\label{fig:speciation}
\end{figure}

\begin{figure}[H] 
\center{\includegraphics[width=0.8\linewidth]{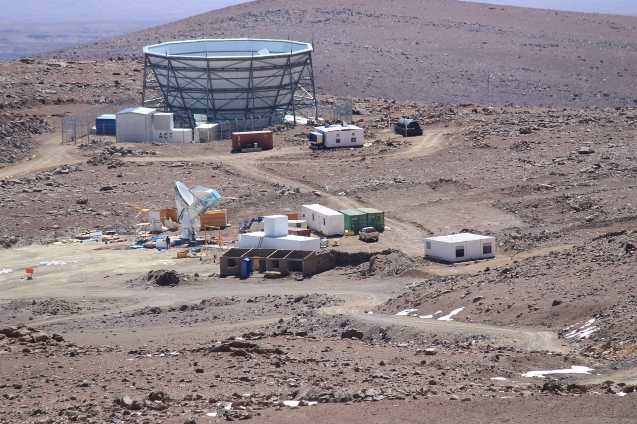}}
\caption{POLARBEAR (foreground) and ACT. (Credit: UCSD Cosmology)}
\label{fig:speciation}
\end{figure}















~

\section{The Future: Commitments and Possibilities} 

In this section, we expand on the future prospects for SMM facilities to which
Canada has access.  For ground-based, single-dish SMM astronomy, Canada
has at present access to the JCMT but Canadian support for its operations is
planned to cease after September 2014.  A new facility, the 25-m diameter CCAT could replace the role of JCMT, but only
as early as 2018.  Moreover, GBT's future is also uncertain, though closure plans
have yet to be articulated.  For interferometric facilities, ALMA's and JVLA's futures
appear more secure.  We reiterate that access to single-dish and interferometric
facilities is crucial in the SMM regime, since they provide respectively the larger-
and smaller-scale views of phenomena that are sometimes only observable at
SMM wavelengths.  Moreover, access to single-dish facilities can provide the
necessary ``ground work" for successful proposals to use the highly oversubscribed
interferometers.  Of note, no other countries with direct access to a significant
shorter wavelength single-dish SMM facility are contemplating its closure as
ALMA comes online.  (Through ESO, the UK and Netherlands will have access
to the 12-m Atacama Pathfinder Experiment (APEX) single-dish telescope.)

For space-based SMM astronomy, the successful Herschel mission is almost
over as the telescope runs out of cryogens.  Though a large and detailed Herschel
dataset will be publicly available worldwide, it will always be limited to the current capabilities
of Herschel.   In this regime, possible successors to Herschel include
``Super BLAST-Pol," a larger aperture follow-up to BLAST-Pol, and SPICA, a
clone of the Herschel design (or possibly larger) with a cooled primary mirror
and instruments optimized for far-infrared/submm observations.




~

\subsection{ALMA} 

ALMA, like the JVLA, is presently finishing its construction and Early Science
phases.  Its future also appears secure, as no other similar or better facility is
planned world-wide.  ALMA plans to keep its capabilities modern through a
healthy development program.  At present, the observatory is considering
several near-term possibilities, including adding Band 1 and Band 2 receivers,
as well as completing the complement of Band 5 receivers (Table 3).  Very long-baseline
SMM interferometry with other facilities around the world is also being explored. 

Though ALMA excels in sensitivity and resolution, it will be limited by its inefficiency
in mapping wide fields.  Hence, single-dish facilities are needed for circumstances
where large-scale data of the SMM universe are needed, or to provide  ``pathfinder"
data for future high resolution data with ALMA.

\bigskip
\begin{figure}[H] 
\center{\includegraphics[width=0.8\linewidth]{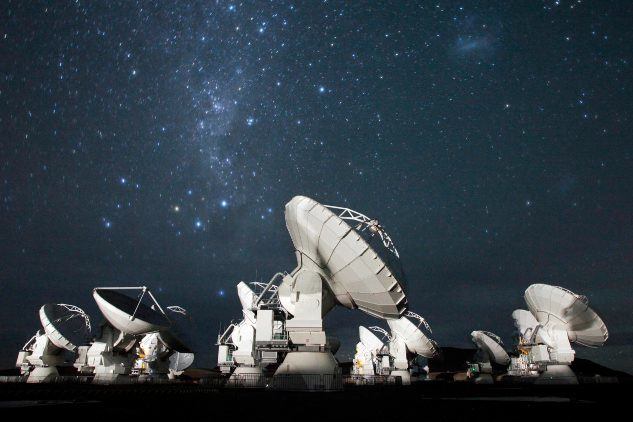}}
\caption{ALMA antennas on the Llano de Chajnantor (Credit: ALMA
(ESO/NAOJ/NRAO))}
\label{fig:speciation}
\end{figure}

\subsection{CCAT}

CCAT is envisaged as an SMM
single-dish facility of 25-m diameter at 5600 m in northern Chile, specifically
located on a peak above the plateau where ALMA lies.  This site is excellent,
having even lower typical precipitable water vapour levels than the adjacent
ALMA site and Mauna Kea.  Also, the CCAT aperture will be $\sim$2.8$\times$
larger than the surface area of the JCMT and $\sim$2$\times$ the diameter of
the 12-m ALMA antennas.  These qualities make
CCAT an attractive possibility for the future of short-wavelength SMM single-dish
astronomy.  

CCAT was identified as one of the top three medium-scale ground-based
observatories in LRP2010, and remains a high priority among the Canadian
community.   It was further identified by the US NRC Astro2010 survey as the
highest priority in a similar medium, ground-based category.  Through a recent
grass-roots effort, eight Canadian universities (McGill, McMaster, Calgary,
Toronto (including CITA, the Department of A\&A and the Dunlap Institute),
UBC, Waterloo, Dalhousie, Western) secured \$550K to join the CCAT 
consortium as a founding member. This membership status provides a
number of tangible benefits such as guaranteed observing time in perpetuity
and an active role in the design of the project through instrument contracts
and seats on the CCAT Board.  The goal of the Canadian consortium is to
have Canada eventually join CCAT as a 25\% partner, mirroring our investment
in the JCMT and ensuring Canadians can make a strong scientific impact in the
field.  The US partners include Cornell, Caltech, University of Colorado, and
AUI and the German partners include the Universit\"at zu K\"oln and Universit\"at
Bonn.  The CCAT operational model will be likely driven more by large surveys
than smaller PI-led projects, though decisions on survey priorities and wider
data access have not yet been made.

\bigskip
\begin{figure}[H] 
\center{\includegraphics[width=0.55\linewidth]{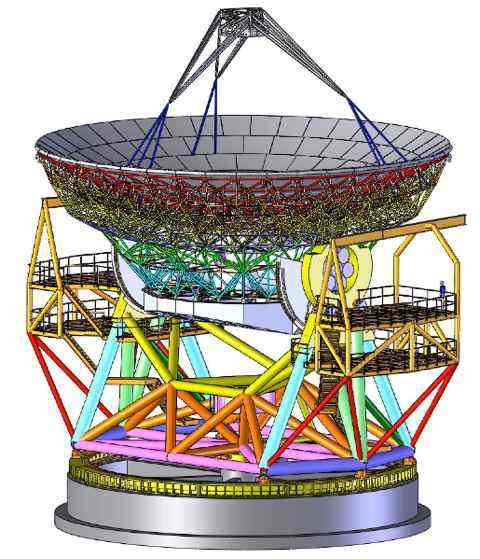}}
\caption{Wireframe model of CCAT. (Credit: CCAT Consortium)}
\label{fig:speciation}
\end{figure}

The designs for the first generation of CCAT instruments are modern and
ambitious, promising very wide-field continuum and spectroscopic imaging.
These instruments will allow wide-field coverage of the southern SMM sky
at sensitivities competitive with those of ALMA.  Four instruments are now
proceeding into an early design phase, after which two are planned to be selected for
first light and the other two for installation during early operations.   These
include:

~

$\bullet${{\bf ~LWCam} - a 
bolometric camera at 5 bands from 750 $\mu$m to 2 mm ($\sim$20$^{\prime}
\times$20$^{\prime}$)};

$\bullet${{\bf ~SWCam} - a bolometric camera at 4 bands from 200 $\mu$m to
670 $\mu$m ($\sim$5$^{\prime}$$\times$5$^{\prime}$)};

$\bullet${{\bf ~X-Spec} - a direct detection wideband survey spectrograph
with instantaneous coverage of 195-520 GHz at $R$ = 400-700
($\sim$20-300 spaxels)}; and

$\bullet${{\bf ~CHAI} - a dual-frequency band heterodyne focal plane array
covering 450 GHz ($\sim$2$^{\prime}$$\times$2$^{\prime}$)
and 830 GHz ($\sim$1$^{\prime}$$\times$1$^{\prime}$).}

~

The  estimate for the cost of CCAT construction (including first instruments) is  \$140M (with first light in $\sim$ 2018). The CCAT telescope itself is presently undergoing a design
and development phase funded by the US NSF that will be completed in 2013, and the road contract has been tendered with construction beginning this year (2013).  
Some private funding for CCAT has been secured by some partners but no
further US or Canadian public funding has been yet identified.  

\subsection{JCMT}
The JCMT partnership in its present form will change in March 2013, when
the Netherlands will withdraw support to JCMT.  UK and Canadian support is
planned to cease at the end of September 2014.  The future of the observatory
after September 2014 is not clear.  A prospectus for new management to take
over operations of JCMT will be released in early 2013.  If no new management
is found, the JCMT will face closure and demolition, with the site returned to its
original condition atop Mauna Kea. Note that JCMT support is beginning to
erode well ahead of September 2014 as experienced staff move to new stable
positions elsewhere.

In Canada, the LRP2000 panel strongly advocated support for JCMT cease
at the end of the tripartite agreement between the UK, Canada, and the Netherlands
to operate the JCMT in 2009, and the funds subsequently freed up be used
for ALMA.  This idea was originally supported by much of Canada's SMM
community as it was based on the key assumption that the JCMT would  
have exhausted its scientific potential as ALMA came online.    This event,
however, has not yet occurred.  For example, plans for SCUBA-2 and the
JCMT Legacy Surveys were not yet formulated during the period when the
LRP2000 panel report was drafted.   Throughout the 2000's SCUBA-2 was
developed, in collaboration with Canada (with funding at the \$5-10M level 
from the Canadian Foundation for Innovation (CFI) made to a university
consortium).   The first SCUBA-2 components usable for observations were
available in early 2010, but science usage of the full instrument did not
begin until fall 2011.  Moreover, the Canadian SMM community at the time
of the LRP2000 was smaller than it is today, more  than a decade later. It is
difficult to assign objective numbers to the increase in community size, but note
that four of the seven authors of this document  were established as part of the
Canadian community after the publication of  the LRP2000 report.

\bigskip
\begin{figure}[H] 
\center{\includegraphics[width=0.55\linewidth]{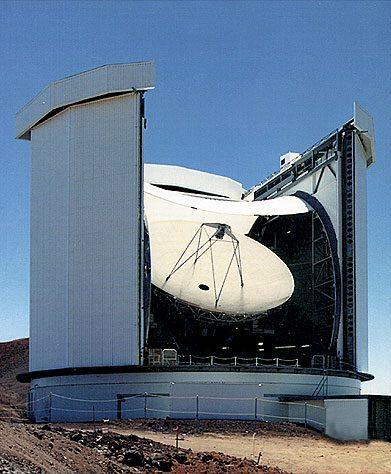}}
\caption{The James Clerk Maxwell Telescope. (Credit: NRC)}
\label{fig:speciation}
\end{figure}

With SCUBA-2 in the picture, the agreement to operate JCMT was
indeed extended beyond 2009 (to the dates described above).
Around the same time, the LRP2010 panel reiterated the earlier LRP2000
report's recommendation of Canadian withdrawal from JCMT, but defined
no specific withdrawal date.  For example, they wrote  ``[w]ithdrawal
from the JCMT is expected shortly but with access to both ALMA and
the EVLA [now JVLA] Canada is now very well positioned at the forefront
of radio/submm astronomy.  Canada is also a member of the CCAT
consortium and is participating in SKA development."  Since the LRP2010
report, the true on-sky performance of SCUBA-2 has been shown to be
excellent and the instrument is still scientifically relevant.  For example,
its mapping speed is $\sim$200$\times$ that of SCUBA at 850 $\mu$m,
and its beam-size at 450 $\mu$m is 4$\times$ smaller than that of
Herschel-SPIRE at 500 $\mu$m.    SCUBA-2 on the JCMT remains the
best facility of its kind in the world. 

Given the present date for withdrawal from JCMT, the Canadian community
will get only three years use of SCUBA-2, its premier instrument, and of
course its remaining instrumentation.  The JCMT Legacy Survey components
were reduced in scale in fall 2011 to $\sim$2 year programs to match the on-sky
performance of SCUBA-2 with the historical weather breakdown at JCMT.  Since
completion assumes the weather holds to statistical norms and no losses of key 
support staff, there is of course associated risk that even the rescoped JLS may
not be fully completed.  (The JLS component teams have prioritized their targets
in this event.)   Moreover, restoring JLS components to their original scopes will
not be possible, meaning that the full scientific potential will not be realized.  Withdrawal from JCMT in September 2014 will also severely
curtail Canadian PI use of SCUBA-2 or HARP, to follow up JLS results, observe
regions of the sky not observed by other facilities (e.g., Herschel), or address any
new developments that demand submm data.  Finally, withdrawal from JCMT will
significantly limit use of the two ancillary instruments developed by Canadians,
POL-2 and FTS-2, that are currently being commissioned and provide unique
data not obtainable at other observatories (i.e., wide-field polarization data and
intermediate spectral resolution mapping, respectively).

Withdrawal from JCMT will also limit Canadians' abilities to build on their legacy
with Herschel.  For example, SCUBA-2 provides information at longer wavelengths
(850 $\mu$m) that Herschel could not provide, at resolutions and sensitivities
similar to those of Herschel at 250 $\mu$m.  In addition, as stated above, SCUBA-2 450 $\mu$m
images have a resolution a factor of $\sim$4 higher than that of Herschel at 500
$\mu$m, significantly reducing possible confusion and source blending seen in
Herschel 500 $\mu$m data.  Though SCUBA-2's data are more filtered spatially
than (Herschel) SPIRE and PACS data, Canadian astronomers have already
begun work combining the complementary JCMT and Herschel continuum data
to great benefit in more accurately determining the column densities of emitting
dust.  Moreover, HARP can be used to produce high spectral resolution data of
molecular lines that provide key kinematic insights about structures seen in either
SCUBA-2 or SPIRE/PACS data (or both).  HARP also has proven to be an ancillary
instrument to SCUBA-2 itself, as HARP data can provide key information about
line emission within the SCUBA-2 850 $\mu$m filter band (mainly from CO) that
may be wrongly attributed to continuum flux.

Terminating Canada's involvement in the JCMT in 2014 will also undermine Canada's investment in ALMA.
While ALMA is a forefront facility, its abilities remain finite and in particular it cannot
realistically compete with facilities that can observe wide sky fields.  JCMT is still
the largest single-dish SMM facility on the planet, and single-dish continuum or line
pathfinder data, such as those from JCMT, give Canadian ALMA proposals a crucial
edge over those of their international peers.  We note again that many countries involved
in ALMA, e.g., Japan, France, Germany, and Spain have {\it not}\/ closed their own
short-wavelength SMM single-dish telescopes as ALMA becomes available, in part
for this reason.  In some cases, the capabilities of these facilities have even been
expanded with new instrumentation or upgrades.  (Though CCAT could replace JCMT
in this role as an ALMA pathfinder, again it will not be available for at least four years
after Canada leaves the JCMT partnership.)  Finally, we note that Canada's use of
ALMA is small relative to the larger partners in that project, limiting our overall scientific
impact as well as the technical and scientific expertise in the SMM regime we can
maintain and foster.  We believe Canada's scientific impact with ALMA will be greatly
augmented through continued access to a large single-dish SMM observatory.

\subsection{GBT and JVLA}

The GBT and JVLA are operated by the National Radio Astronomy Observatory
(NRAO) in the US through funding from the NSF to Associated Universities, Inc.
(AUI).  NRAO also operates ALMA on behalf of North America and the 
radio-frequency Very Large Baseline Array.  Canada has secured access to the
NRAO facilities through NAPRA.

\bigskip
\begin{figure}[H] 
\center{\includegraphics[width=0.8\linewidth]{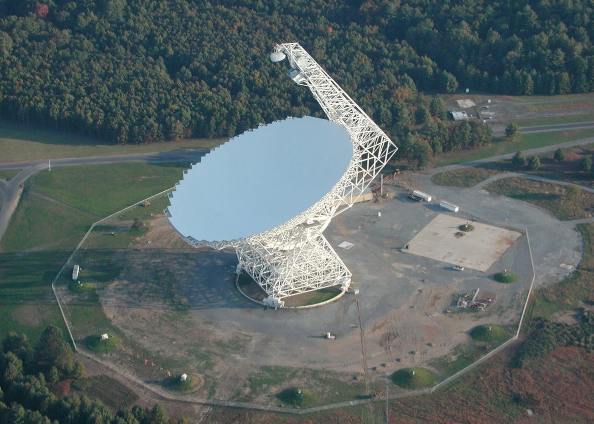}}
\caption{The Robert C. Byrd Green Bank Telescope. (Image courtesy of NRAO/AUI)}
\label{fig:speciation}
\end{figure}

The GBT's future is uncertain given the August 2012 recommendation by the
NSF Portfolio Review Committee that NSF divest itself from its commitment to
operate the  GBT, so that funding can be freed up for facilities to address new
astronomical priorities.  As of this writing, it is not yet known how NSF, AUI,
or NRAO will act on this recommendation, e.g., modifying (reducing) the
present operational model for GBT, putting forth the facility for management by
new (national or international) partners, or closing it outright.  Loss of the GBT
(or even loss of open access to it) would be a serious blow to single-dish
long-wavelength SMM (and radio) astronomy.  The GBT's surface and imaging
capabilities exceed those of its closest competitor, the Effelsberg 100-m 
Telescope in Germany.  Some members of the US (and Canadian) community
are actively campaigning to keep open access to the GBT.

The JVLA's future is more secure, as its retrofit and Early Science phase have
just been completed.  The JVLA is the world's foremost radio and long-wavelength
SMM facility.  As with ALMA, the ability to see the large-scale picture is lost with
the JVLA but can be restored with data from single-dish facilities.  Hence, the 
loss of the GBT to general use would also impact the scientific potential of the
JVLA.

\bigskip
\begin{figure}[H] 
\center{\includegraphics[width=0.8\linewidth]{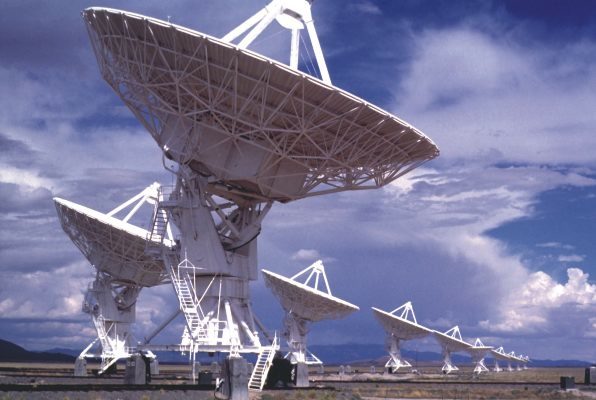}}
\caption{The Karl G. Jansky Very Large Array. (Image courtesy of NRAO/AUI)}
\label{fig:speciation}
\end{figure}

Future developments of the JVLA have not been openly discussed within the 
US community.  ALMA Band 1 receivers on ALMA could exceed the performance
of the present 40-50 GHz (Q-band) receivers on JVLA.  Further along, the idea
of providing mid-scale baselines for the VLBA through a
network of single-dish JVLA-like telescopes throughout New Mexico, linking the
JVLA with the VLBA, could be resuscitated.  (Such a development depends on the future of the VLBA
and is beyond the scope of this document.)  Eventually, JVLA's capabilities at
radio frequencies $<$10 GHz will be eclipsed by those of the SKA.  

\subsection{SPICA}

The Space Infrared Telescope for Cosmology and Astrophysics (SPICA)
is a planned mission optimized for mid- and far-infrared astronomy with a
cryogenically cooled telescope.  The baseline telescope is a clone of the 3.5-m
Herschel telescope, though other configurations are being considered.  To
reduce mass, SPICA will be launched at ambient temperature and cooled down
in orbit by onboard mechanical coolers with an efficient radiative cooling system.
It has been proposed as a Japanese-led JAXA-ESA mission together with
extensive international collaboration including Canada, the US, and South
Korea.  As of 2013, the proposed launch date for SPICA is 2022. 

\bigskip
\begin{figure}[H] 
\center{\includegraphics[width=0.8\linewidth]{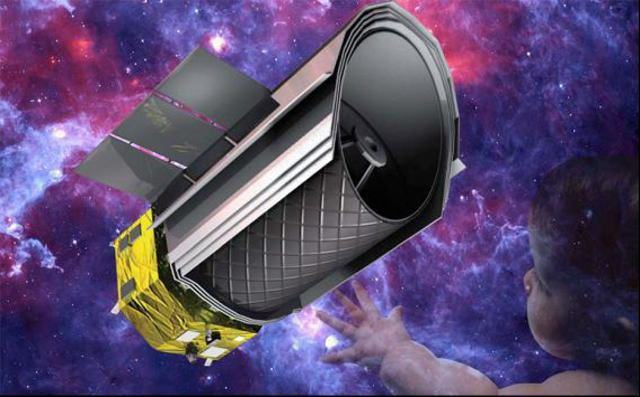}}
\caption{The Space Infrared Telescope for Cosmology and Astrophysics
(SPICA) concept.  Space ghost baby likely not part of final design. (Credit: JAXA)}
\label{fig:speciation}
\end{figure}

Though optimized for wavelengths mostly shorter than the SMM range, SPICA
is of great interest to Canadian astronomers given the success and legacy of 
the Herschel mission.  Through SPICA's instruments, data will be acquired that
are highly complementary to Herschel's, bridging the presently under-sampled
wavelengths between the infrared and SMM regimes also highly impacted by
atmospheric absorption.  At wavelengths shorter than those covered by Herschel,
SPICA data will naturally be of higher resolution than Herschel's.  Moreover, its
cooled primary will allow for significant improvements in sensitivity over Herschel. 
The three baseline SPICA instruments include:

~

$\bullet$ an unnamed mid-infrared (MIR) coronagraph (3.5-27 $\mu$m) with
photometric and spectral capabilities ($R$ $\approx$ 200);

$\bullet$ an unnamed MIR wide-field camera and high-resolution spectrometer
($R$ $\approx$ 3 $\times$ 10$^{4}$); and

$\bullet${\bf ~SAFARI}  - a far-infrared (30-210 $\mu$m) imaging spectrometer.

~ 

Canadian astronomers are currently involved in a detector test facility and
simulator for SAFARI, primarily through D. Naylor's group at U. Lethbridge.
A fourth instrument, BLISS, a low-resolution ($R$ $\approx$ 500) spectrograph
for extragalactic surveys is under design in the US and Canada in consultation
with Japan, but based primarily on NASA funding.  This instrument is currently
exploring several designs and will focus on one of a grating spectrograph, an
imaging FTS, or a filter-bank architecture in the coming years. 


\subsection{Focused PI Experiments}

\subsubsection{The E and B Experiment}
The E and B experiment (EBEX) is a mm-wavelength polarization-sensitive 1.5-m
diameter telescope that will be flown by NASA to about 100,000 feet aboard a
stratospheric balloon.  Detection of polarization from inflationary gravity waves is
one of the main science goals of EBEX.  EBEX will measure the polarization of
the CMB to provide a glimpse of the universe at its very earliest stages.  The 
focal plane of EBEX will be instrumented with 1400 Transition Edge Sensor (TES)
detectors, read out with McGill's Digital Frequency Domain Multiplexer electronics.
Polarization sensitivity is achieved with a half-wave plate and polarizing grid.  It will
flown first on a one day test flight above Texas, before traveling to Antarctica for a
$\sim$30 day long duration flight around the South Pole.

\subsubsection{Spider}
Spider is a balloon-borne telescope designed to detect the imprint of gravitational
waves released in the the first tiny fraction of a second of the universe and thereby
imprinted on the CMB. By doing this, Spider will provide insight on the extremely early
Universe, and provide a crucial test for models of the early inflation of the Universe.
Spider has polarization-sensitive detectors at 100 GHz (3 mm), 150 GHz (2 mm) and
220 GHz (1.4 mm). The first launch expected in summer of 2013.

\subsubsection{Super\ BLAST-Pol}
Super\ BLAST-Pol is an upgrade of the existing BLAST-Pol telescope (the Balloon-borne Large Aperture Sub-mm Telescope for Polarimetry), which will map polarized dust emission at 250, 350 and 500 $\mu m$ with a resolution of 22$''$ at 250 $\mu m$.  The telescope will utilize a 2.5 meter aluminum primary mirror, a 28-day hold time cryostat, and a 1,000 detector focal plane array using MKIDs detector technology, which will give Super\ BLAST-Pol a mapping speed $>$10$\times$ that of BLAST-Pol.    The project has funding from NASA, and is being built by the University of Pennsylvania, Northwestern University, Arizona State, NIST, the University of Toronto, Cardiff University and the University of British Columbia.   A first science flight is planned for December 2016 from Antarctica: 25$\%$ of the science time will be available for Òshared riskÓ observing to the astronomy community, making Super\ BLAST-Pol the first balloon-borne telescope to operate as an observatory.

\subsubsection{Event Horizon Telescope}
The Event Horizon Telescope (EHT) is an international consortium to link numerous SMM facilities around
the globe into a single very long baseline interferometer.   The goal of the EHT is to observe directly
230-450 GHz emission from the immediate environments of supermassive black holes in the centres
of our Galaxy and the nearby elliptical galaxy M87.  Such observations will probe accretion and jet
formation in these highly unusual sites, and test general relativity.  Among its many nodes, the EHT will
include the JCMT and ALMA.  This ambitious project, including participation from the University of Waterloo
and the Perimeter Institute of Theoretical Physics, is expected to take a decade to complete due to
the need to develop and deploy highly stable frequency standards, new submillimetre dual-polarization
receivers, and wide band width VLBI backends and recorders.

\subsection{Synergies with Other Facilities}

In this document, we have focused entirely on SMM facilities to which Canada has
access.  Through its long engagement in JCMT, Canada has built up a considerable
community of astronomers who use SMM wavelengths to investigate astrophysical
phenomena.  Of course, the  SMM range is only a small part of the wider electromagnetic 
spectrum.  How do SMM facilities relate to those given high priority (e.g., in LRP2010)
at other wavelengths?

Optical/infrared telescopes like TMT or Euclid are important for studying the warmer
thermal emission from stars (and planets) and galaxies.  SMM facilities are used to 
explore the colder aspects of nature, typically the cold interstellar media of this Galaxy 
and others.  Beyond extinction and interstellar absorption, the cold ISM is generally
not probed at optical/infrared wavelengths.  The ISM is important, e.g., accounting for
mass in spiral galaxies similar to those seen in their stellar components.  

Similarly, advanced radio telescopes like CHIME and the SKA are important for
probing both the ionized components of galaxies through thermal free-free emission
and the warmer atomic aspect of the ISM within galaxies through observations of HI.
Moreover, such facilities are critical for exploring the signatures of non-thermal
processes in the universe, e.g., synchrotron emission.  Few molecules strongly emit at
the wavelengths generally considered to be ``radio" in nature, and thermal dust
emission becomes too weak at wavelengths $>$10 mm to be detected.  

Hence, the SMM regime is in many ways a unique realm where specialized 
equipment both on the ground and in space are needed to unlock its secrets.
Though the optical/infrared and radio facilities prioritized in LRP2010 are very
important, attention must be paid to maintain Canada's leadership in the SMM
regime through continued engagement in such facilities on many levels, i.e., 
beyond just that of ALMA.

\newpage

\section{A Plan for the Next Decade}

\subsection{Recommendations} We recommend that Canada retain its standing in SMM
astronomy by building on its existing strengths and acting on its present and future opportunities.
To do so, Canada must retain access to the unique strengths of a single-dish telescope at a
level sufficient to maintain and foster scientific and technical expertise.  
In particular, we present the following recommendations:

~

$\bullet$ Canada must continue to encourage and maintain active participation in ALMA science proposals and in ALMA Development Projects. As a member of the North American ALMA ARC with full access to 33.75\% of ALMA time, Canadian PIs need only have strong science cases to achieve high rates of allocated time on ALMA. The results of the first two cycles of ALMA allocations show that Canada is already competing well, obtaining time in excess of our basic monetary contributions. ALMA Development is an ongoing process of furthering the capabilities of ALMA, and Canada has been engaged in Band 1 receiver development for several years. This and other development opportunities should be pursued so that Canada's science objectives for ALMA are paralleled by hardware and software contributions from Canada as well.

~

$\bullet$ Canada must continue engagement within the JCMT with a smooth transition to CCAT.
An extension of our involvement by a minimum of three years (to the end of 2017) will allow for the
completion of the Legacy Surveys to their intended (peer-reviewed) level.    These surveys remain
highly relevant  with no comparable facility capable of their execution.  Such a timescale
is also concurrent  with the expected availability of CCAT of 2018.   Moreover, the extensive JLS data
can continue to involve the community for several years as CCAT ramps up its capabilities.   
 
Continued support for JCMT would
have the additional benefit of enabling Canadians to follow up discoveries
made by the JLS program and from Herschel data. These new observations would be made
using the JCMT's present suite of 
unique instruments and will 
 realize the scientific potential of the two instruments POL-2 and FTS-2 built
with Canadian funding.  Access to JCMT will also give Canadians
an important edge when devising programs for  highly competitive ALMA, for which few pathfinder
instruments exist. Finally, continued use of the JCMT will  provide a training ground and test-bed
facility for CCAT for a growing cohort of involved Canadian students and postdocs.

Canadian support for JCMT is presently at the \$1M/annum level.  Though the
future of the partnership of JCMT remains unclear, Canada should be poised to
continue engagement at its present level of 25\% within whatever consortium may
form to take over management of the JCMT.  A mechanism for continued Canadian
involvement in JCMT must be immediately found.


~

$\bullet$ Canada must move forward with its engagement in CCAT.  The minimum level of participation
set by the partnership is 10\%, but we recommend 25\% if Canada is to have a strong voice in the
collaboration, a meaningful scientific impact, and the opportunity to maintain and foster expertise.  Given
its size and location, CCAT will be clearly superior to JCMT and we do not foresee a need for both facilities.
Thus, CCAT first-light represents a hard
upper-limit on continued Canadian involvement  in the JCMT. CCAT will allow Canada to continue building
on its heritage in the SMM regime obtained  through its engagement in JCMT, and allow for continued
advantage for using highly over-subscribed facilities like ALMA.

~

$\bullet$ The JVLA is an impressive facility coming into its own thanks in large part to the Canadian-made
WIDAR correlator.  We recommend Canadians take advantage of its unique and powerful capabilities in the longer-wavelength millimetre regime.  Regarding the GBT, the recent recommendation to NSF that it divest itself from its operations is worrisome in that Canadian access to future single-dish long-wavelength SMM data are threatened.  GBT data are impressive on their own, or notably can be used to great effect in combination with JVLA data.  The GBT is the top of its class, and no other facility of its type is planned on any timescale.  Though Canada has access to GBT or JVLA through the NAPRA agreement, it does not directly fund either facility.  How NRAO will proceed on the recommendation to divest from GBT is unknown at this time.  We recommend, however, that Canadians stay abreast of potential changes in its operations over the next decade and at the very least be prepared to advocate for continued access.

~

$\bullet$ Canada, specifically the Canadian Space Agency, should leverage its
previous successful investment with Herschel to fund Canadian engagement in
SPICA.  As with Herschel, the CSA should consider funding students and postdoctoral
fellows to allow Canadians to exploit fully its investment before the data become widely
public.

~

$\bullet$ Though general purpose observatories are extremely useful, sometimes
very focused PI-based experiments can provide the answers to pressing problems
that arise after larger, open-access observatories and their instruments are built.  These
nimble experiments can also make great strides with relatively small cost, particularly
those based on balloons.  Hence, we recommend ongoing funding to  focused PI-experiments,
like EBEX, Spider, Super BLAST-Pol, POLARBEAR, and ACT through the Canadian
Space Agency, CFI, and NSERC.

\subsection{Funding Considerations}

\subsubsection{JCMT}

At present, Canada's involvement in JCMT is provided by the National Research
Council of Canada (NRC).  NRC is mandated to  support  Canada's
off-shore astronomy facilities (e.g., Gemini and CFHT) and has begun to fund
Canadian participation in ALMA.  For Canada to remain engaged in JCMT, 
the most obvious solution is for NRC to provide new support and continue to manage
its operations. Alternatively the Canadian  astronomy
community must find funding from other national sources.  Such a mechanism is
not yet clear, but perhaps one where funding from CFI or NSERC managed by
ACURA is possible, bridging to future management of CCAT.

Current Canadian operating costs of the JCMT are \$1.17M per year, and
last year's total operating cost of the JCMT was \$4.7M (US). Even 
considering a stripped down mode of operation for the JCMT of \$2M/yr,
with Canada continuing to contribute 25\% of the costs, then the
funding required would be \$500k/yr. While not a trivial amount of money it is also not 
enormous, when one considers the size of the JCMT user base, its scientific impact,
and the investment to date. 
Still, at present there are no appropriate programs available through NSERC or CFI. 

The initial \$500k which secured the Canadian consortium partnership in CCAT was raised through one-time requests to 
individual universities and institutes, with each contributing $\sim$\$50k.  While this success confirms that this level of funding is feasible at
the university level, this approach is unlikely to possible for continued  JCMT funding, as it is not a new initiative.

\subsubsection{CCAT}

The Canadian CCAT Consortium plans to submit a proposal for funding to the Canadian Foundation
for Innovation (CFI) this year. 
The CCAT consortium, including Canadian universities and institutes, are actively exploring the potential for private donations. The required funds for construction (\$35M) are within reach of a single, or a small number of private donors.
The ongoing costs of operation of CCAT, however, are unlikely to be funded through these mechanisms.

We note that the expertise of NRC in managing telescopes and providing archival resources makes it a natural
partner for the CCAT project.   We recommend
that the management and operations structure of CCAT be finalized soon, with
some clear guidance from Canadian funding agencies for continued and stable
means to fund CCAT operations.

\subsubsection{SPICA and PI Projects}

The Canadian Space Agency (CSA) is the primary means of funding
involvement in new space missions such as SPICA. They also support
various smaller programs related to balloon missions, which are relevant
and useful for the PI-types of projects in particular. Over the past decade, CSA has
provided significant funding resources, including roughly \$100M for
our participation in JWST, Herschel, and Planck, as well as \$1-2M of
funding in support of data analysis associated with Herschel, Planck,
FUSE, and other operating missions to individual university
researchers. CSA thus has the potential in the long run to provide
major new funding for space- and balloon-based submillimetre
facilities. 

The CSA's budget is currently very tight and informal feedback
suggests they have no money for new astronomy projects at this
time. Budget situations can change dramatically, however, with new
directions and funding coming from the federal budget process. The community
at large will need to continue to put pressure on the CSA and at the
political level to lobby for additional funding for science programs
in the CSA's budget.

\vskip 1.5cm
\noindent{\it Acknowledgements:}  The authors thank Laura Fissel for providing information on Super BLAST-Pol, David Naylor for providing information on SPICA, Gerald Moriarty-Scheiven for providing the ALMA user numbers, Dennis Crabtree for providing the JCMT, CFHT, and Gemini user statistics, and Tom Phillips for providing an updated version of
Figure 1.  We also thank Matthijs van der Wiel and Gary Davis for pointing out small errors in a previous version of this document.






 


\end{document}